\documentclass[aps,prb,10pt,twocolumn,footinbib,superscriptaddress,floatfix]{revtex4-2}

\usepackage{CJK}
\usepackage{times}
\usepackage{multirow}
\usepackage[dvipsnames]{xcolor}
\usepackage{amsmath}
\usepackage{amsthm, mathrsfs}
\usepackage{amssymb}
\usepackage{amsbsy}
\usepackage{enumitem}
\usepackage{bm}
\usepackage{wasysym}
\usepackage[english]{babel}
\usepackage[T1]{fontenc}
\usepackage[utf8]{inputenc} 
\usepackage{graphicx}
\usepackage[colorlinks,bookmarks=false,citecolor=blue,linkcolor=red,urlcolor=blue]{hyperref}
\usepackage{pstricks}
\usepackage{rotating}			       
\usepackage{tabularx,hhline}	
\usepackage[caption=false]{subfig}		
\usepackage[normalem]{ulem}

\newcolumntype{P}[1]{>{\centering\arraybackslash}p{#1}}

\begin{document}

\begin{CJK*}{UTF8}{gbsn} 
\title{
Conformal Boundary as Holographic Dual to the Hyperbolic Fracton Model
}


\author{Alejo Costa Duran}
\affiliation{
Instituto de F\'isica de L\'iquidos y Sistemas Biol\'ogicos (IFLySiB), CONICET and Universidad Nacional de La Plata,  La Plata, Argentina}
\affiliation{CCT CONICET La Plata, Consejo Nacional de Investigaciones Cient\'\i{}ficas y T\'ecnicas, Argentina}

\author{Mauricio Sturla}
\affiliation{
Instituto de F\'isica de L\'iquidos y Sistemas Biol\'ogicos (IFLySiB), CONICET and Universidad Nacional de La Plata,  La Plata, Argentina}
\affiliation{CCT CONICET La Plata, Consejo Nacional de Investigaciones Cient\'\i{}ficas y T\'ecnicas, Argentina}

\author{Ludovic D.C. Jaubert}
\affiliation{CNRS, Universit\'e de Bordeaux, LOMA, UMR 5798, 33400 Talence, France} 

\author{Han Yan (闫寒)}
\affiliation{Institute for Solid State Physics, 
The University of Tokyo, Kashiwa, Chiba 277-8581, Japan}
\email{hanyan@issp.u-tokyo.ac.jp}
\date{\today}
\begin{abstract}
In addition to describing our universe, gravitational theories profoundly inspire the study of emergent properties of exotic phases of matter.
While the Anti-de Sitter/conformal field theory (AdS/CFT) correspondence is one of the most celebrated examples, the field of fractonic matter -- driven in part by gapless phases resembling linearized gravity -- has also seen rapid developments. 
Despite the deep implications of both areas, connections between them remain sparse, primarily due to the difficulty in constructing explicit models that encapsulate both fields' essential features.
Here we demonstrate the efficacy of the recently proposed Hyperbolic Fracton Model as a concrete model for AdS/CFT duality. Using explicit numerical and analytical calculations on the discrete hyperbolic lattice, we show that the boundary state exhibits conformal field theory properties. Our main result is that bulk defects induce an emergent temperature for the boundary state,  proportional to the defect perimeter, in quantitative agreement with the expected behaviour of a black hole in AdS spacetime. The Hyperbolic Fracton Model thus emerges as a unique lattice model of  holographic principle equipped with a well-defined \textit{bulk} Hamiltonian, and offers a promising gateway for studying a wide range of holographic phenomena. 
\end{abstract}
\maketitle
\end{CJK*} 

\section{Introduction}

The Anti-de Sitter/conformal field theory (AdS/CFT) correspondence~\cite{Maldacena1999,Gubser1998PLB,Witten1998} offers deep insights into the relationship between gravitational theories in AdS spacetime and CFTs on the boundary. This duality has provided a powerful framework for understanding quantum gravity, black hole (BH) thermodynamics, and strongly coupled quantum systems, revolutionizing our approach to fundamental problems in high-energy physics, condensed matter and beyond~\cite{Susskind1995arXiv,hooft2009arxiv,Zaanen2015_ADSCMTbook,Hartnoll:2018xxg,breuckmann2020critical}.

But recent years have seen the apparition of a novel framework where condensed matter meets gravity. Fractons are quasi-particles with intrinsic restricted mobilities \cite{Nandkishoreannurev,pretko2020fractonReview}. In their gapless form,  lattice gauge theories of fractons were developed in a effort to mimic linearized gravity \cite{Pretko2017PhysRevD,Yan20a}; while gapped fracton order arises from the search of self-correcting quantum   codes  and gauging subsystem symmetries\cite{Haah2011,Vijay2015}.

Defined on a surface with  negative curvature, the hyperbolic fracton model (HFM)~\cite{Yan2019PhysRevBfracton1,Yan2019PhysRevBfracton2} bridges the gap between fractonic matter with subsystem symmetries and various AdS/CFT models including holographic tensor networks~\cite{Swingle2012,Pastawski2015,Almheiri2015,Yang2016,alex2019majorana,Jahneaaw0092} and bit-threads models~\cite{Freedman2017CMaPh}, thus contributing to the challenging task of constructing explicit models to understand  AdS/CFT.
Together, these discoveries place fracton phases at the crossroad between topological order, exotic quantum phases and dynamics, quantum information, and quantum gravity.
 
\begin{figure}[t!]
    \centering
    \includegraphics[width = 0.9\linewidth]{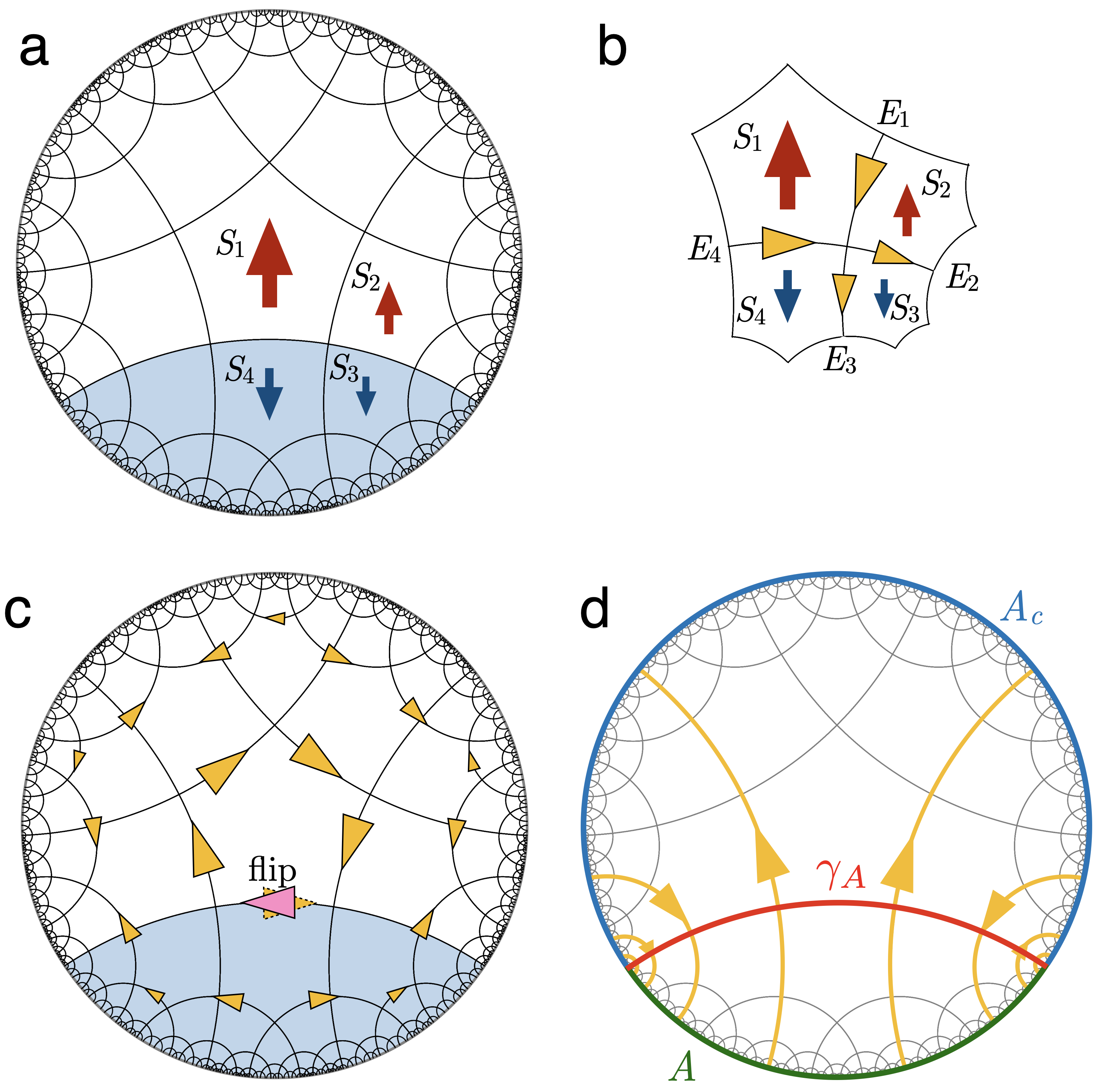}
    \caption{
    (a) The hyperbolic fracton model on the $(5,4)$ tesselation of the hyperbolic plane, represented as the Poincar\'e disk. Four spins are shown in the lattice to illustrate the Hamiltonian [Eq.~\eqref{EQN_HFM_Hamiltonian}]. An example of subsystem symmetry is to flip all spins in the light blue region (a geodesic wedge). 
    (b) Definition of the dual vertex model. The yellow arrows are the degrees of freedom of the vertex model. The definition is given in Eq.~\eqref{EQN_def_of_E}.
    (c) The ground states in the vertex model are those such that the arrows on the same geodesic are aligned, and different geodesics are independent. Subsystem symmetry acts as flipping the arrow on the geodesic of the wedge, colored in pink. 
    (d) The entanglement entropy between boundary region $A$ (green) and its complement $A_c$ (blue) is measured by the number of lattice geodesics connecting the two sides, highlighted by yellow color. This quantity is equivalent to the length of minimal covering curve $\gamma_A$ separating  $A$  and $A_c$, colored in red.
    }
    \label{fig:FIG_1_lattice}
\end{figure}

In this paper, our motivation is to advance this network of connections between gravity, holography and strongly correlated systems. Using the HFM as our springboard, we analyze the exact nature of its boundary states. We first show,  that boundary states exhibits the characteristics of a CFT in its entanglement entropy and correlation. Then our main result comes from introducing lattice defects to the hyperbolic lattice. These bulk defects can be seen as an open boundary puncture on the lattice, and act as analogs of black holes in the bulk. They generate an emergent temperature for boundary CFT directly proportional to the perimeter of the defect, in agreement with the Ryu-Takayanagi formula at finite temperature \cite{Ryu2006PhysRevLett,Ryu2006JHEP}. The HFM thus provides a well-defined \textit{bulk} Hamiltonian and  quantitatively emulates black-hole physics in the context of AdS/CFT.

\section{Results}

\subsection{The hyperbolic fracton model (HFM)} 

In its simplest form, the HFM is defined on a (5,4) tessellation of the two-dimensional hyperbolic plane, forming a lattice composed of pentagons, with four pentagons meeting at each vertex [Fig.~\ref{fig:FIG_1_lattice}(a)]. In the Poincar\'e disk representation, every arc in the lattice corresponds to a geodesic of the hyperbolic plane. Each pentagon labeled $i$ in the lattice hosts an Ising spin $S_i = \pm 1$. The Hamiltonian of the model is given by: 
\begin{equation}
\label{EQN_HFM_Hamiltonian}
    H = -\sum_{v}S_{v,1}S_{v,2}S_{v,3}S_{v,4} ,
\end{equation}
where $v$ runs over all vertices, and $i$ denotes the four spins around each vertex [Fig.~\ref{fig:FIG_1_lattice}(a)].
We also introduce a cut-off in the lattice to define its boundary: boundary vertices with fewer than four surrounding spins do not contribute any terms to the Hamiltonian. The boundary sites are then all the pentagons next to the boundary, and all other sites inside are the bulk sites.

The ground states of this Hamiltonian can be explicitly constructed by utilizing its subsystem symmetries: by selecting any geodesic on the lattice and flipping all the spins on one side, the Hamiltonian remains invariant.
The light blue wedge in Fig.~\ref{fig:FIG_1_lattice}(a) shows an example. 
Starting from an initial ground state where all spins point up, applying these subsystem symmetries of different combination of  geodesics can generate  all the other ground states.

Flipping a single spin generates five excitations at the five vertices of the corresponding pentagon. These excitations are classical analogs of fractons \cite{ChamonPhysRevLett.94.040402,YoshidaPhysRevB.88.125122,BRAVYI2011839,Haah2011,Vijay2015,Pretko2017a}. Notably, a single excitation cannot be moved across the lattice by flipping only a local patch of spins.

\begin{figure}[t]
\centering
\subfloat[]{\includegraphics[width=\linewidth]{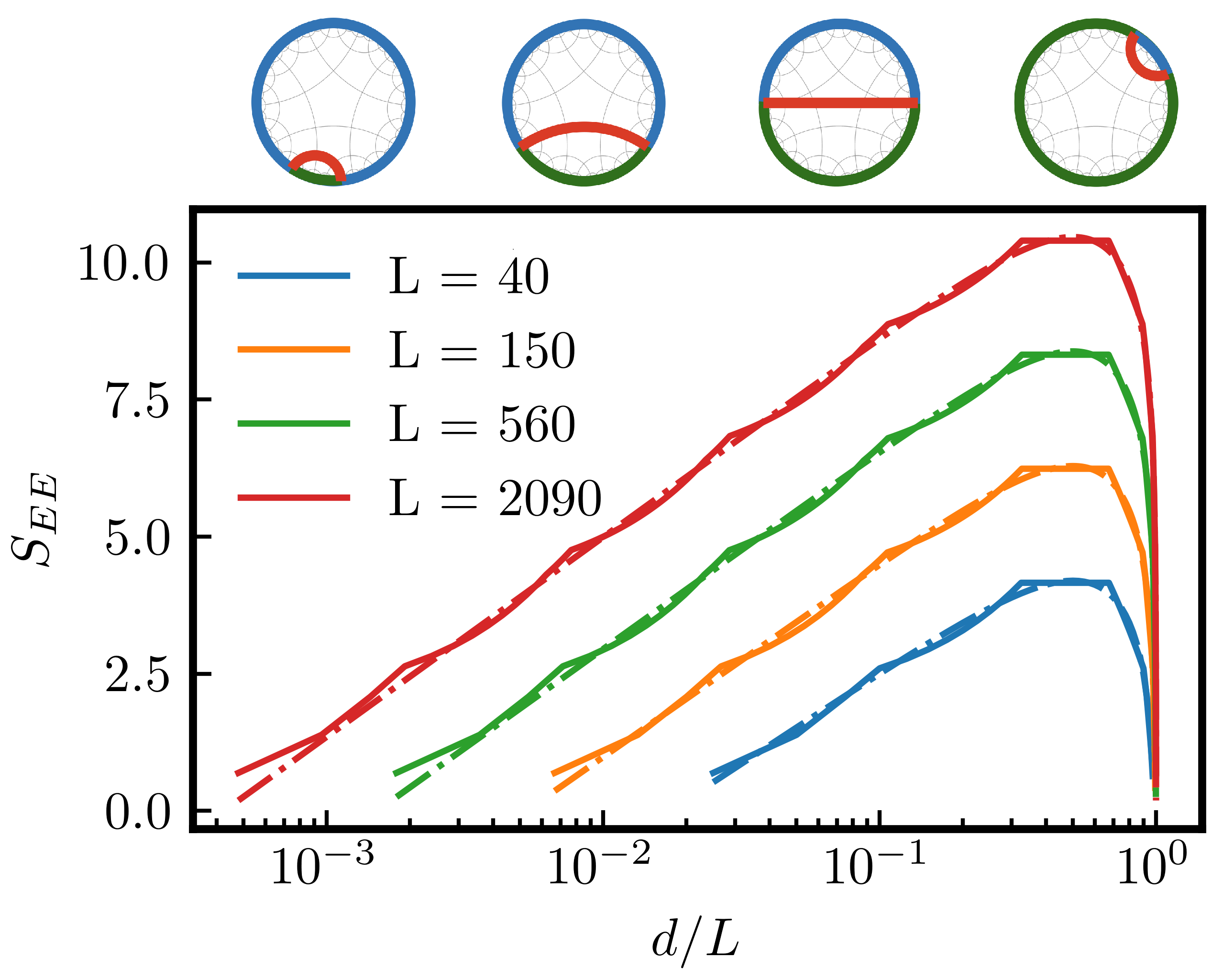}\label{fig:zeroTempEntropies}}\\
\subfloat[]{\includegraphics[width=0.8\linewidth]{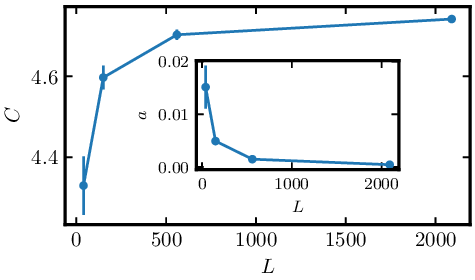}\label{fig:CentralCharges}}
\caption{
(a) Entanglement entropies as a function of boundary segment size $A$ for different lattice sizes $L$. In dashed lines the theoretical prediction from the Calabrese-Cardy equation \eqref{eq:Calabrese-Cardy-Zero}. The upper row panels schematically show the evolution of the minimal covering curve separating $A$ (green contour) and $A_c$ (blue contour).
(b) Central charge $C$ as a function of the system size $L$. The values for both the central charge and the parameter $a$ shown in the inset were obtained through a least squares fit of panel (a).
}
\end{figure}

The HFM is equivalent to a vertex model, a useful mapping providing a pictorial description of the underlying physics. We define binary arrows, denoted as $E_e$, on the edges of the lattice as: 
\begin{equation}
\label{EQN_def_of_E}
E_e = \eta_g \, S_{e,l} \, S_{e,r},      
\end{equation}
where $(e,l)$ and $(e,r)$ refer to the two pentagons adjacent to the edge $e$ on geodesics $g$, and $\eta_g=\pm 1$ is a gauge choice orienting each geodesics $g$. For example, in Fig.~\ref{fig:FIG_1_lattice}(b), $E_1 = S_1 S_2$. The ground state condition at a vertex is given by:
\begin{equation}
S_1 S_2 S_3 S_4 = 1 \Leftrightarrow E_1 = E_3 \ \&   \ E_2 = E_4  .
\end{equation}
In the vertex model, the ground states are characterized by the following conditions:
(1) all arrows $E_{e\in g}$ on the same geodesic $g$ must align in the same direction; 
(2)  each geodesic can select its direction $E_{e\in g}$ independently from the others [Fig.~\ref{fig:FIG_1_lattice}(c)]. 
In this dual vertex model, the subsystem symmetry along a geodesic wedge flips all the arrows on that geodesic, providing a clear and intuitive way to visualize the  ground states and compute the correlation and entanglement quantities.
We note that the same picture has arisen from holographic tensor-networks~\cite{alex2019majorana,Jahneaaw0092}, bit-thread models~\cite{Freedman2017CMaPh}, and higher-rank gauge theory in curved space~\cite{Yan2020PhysRevBGeodesicString} too.

\subsection{Conformal field theory on the boundary}

First, let us show that, on a pristine hyperbolic lattice, the boundary states of the HFM possess the characteristics of a conformal field theory at zero temperature. To define the boundary state in this classical model, we first construct a mixed  state density matrix by taking an equal-probability sum over all ground states, $\rho_{\text{g.s.}} = \frac{1}{\mathcal{N}}\sum_{i \in \text{g.s.}}|\phi_i\rangle \langle\phi_i|$ where $\mathcal{N}$ is a normalizing factor --- essentially, this represents the ground state ensemble at zero temperature. By tracing out the bulk degrees of freedom, we derive the boundary state reduced density matrix
\begin{equation}
\rho_{\text{bdry}} = \mathrm{Tr}_{S_i \in \text{bulk }  }\;\rho_{\text{g.s.}},
\end{equation}
which can be understood as an equal-probability sum of all boundary configurations associated with the ground states.
Note that the subsystem symmetry always acts on part of the boundary sites. 
Hence every product state on the boundary in this density matrix has the same probability.
For a boundary segment $A$ of size $d$, and its complement $A_c$, we have the entanglement entropy defined as  
\begin{equation}
S_{EE}(A) = - \mathrm{Tr}\left[\rho_{\text{bdry},A} \ln \rho_{\text{bdry},A} \right], 
\end{equation}
where $\rho_{\text{bdry},A} $ is the reduced denstiy matrix on $A$.
Note that each state can be written as a product of independent lattice geodesics making different choices of its directions, the entanglement entropy can be written exactly as  
\begin{equation}
S_{EE}(A) = N_{A,A_c}\ln 2,
\end{equation}
where $N_{A,A_c}$ is the number of geodesics connecting $A$ and its complement $A_c$~\cite{Yan2019PhysRevBfracton1}.
We note that, the entanglement entropy in this case is identical to the mutual information between $A$ and $A_c$, defined as  $I(A,A_c) = S_A + S_{A_c} - S_{A \cup A_c}$, where $S_A$ is the von Neumann entropy of the reduced density matrix on $A$~\cite{von1955Bookmathematical}. 
The outcome corresponds to the length of the minimal covering curve $\gamma_A$ separating $A$ from $A_c$ [Fig.~\ref{fig:FIG_1_lattice}(d)], which is analogous to the area of minimal covering surface in AdS/CFT \cite{Ryu2006PhysRevLett}.  
The $S_{EE}$, as a function of boundary size $d$ and averaged over all choices of connected $A$, is numerically computed, whose details and the code used can be found in \cite{FractonGitHub}.
We find that the numerical results of Fig.~\ref{fig:zeroTempEntropies} closely follow the Calabrese-Cardy formula for the entanglement entropy of a periodic $(1+1)$-dimensional CFT of finite length $L$ at zero temperature \cite{calabrese2004entanglement}:
\begin{equation}
S_{EE} = \frac{c}{3}\ln \left( \frac{L}{\pi a} \sin \left( \frac{\pi d}{L} \right) \right).
\label{eq:Calabrese-Cardy-Zero}
\end{equation}
The central charge $c$ of the CFT and free parameter $a$ are extracted from the fit to Eq.~(\ref{eq:Calabrese-Cardy-Zero}) and converge to $c = 6.840(9)\times \ln 2$ and $a = 0$ in the thermodynamic limit [Fig. \ref{fig:CentralCharges}], which  agrees well with the perfect tensor-networks  result $9\ln2/\ln(\sqrt{3}+2)$ on the same tessellation~\cite{Jahn2020PhysRevA}. The  result  is consistent with $a$ playing the role of an effective lattice constant, as we approach the continuum limit \cite{calabrese2004entanglement}.

\begin{figure}
    \centering
    \includegraphics[width = \linewidth]{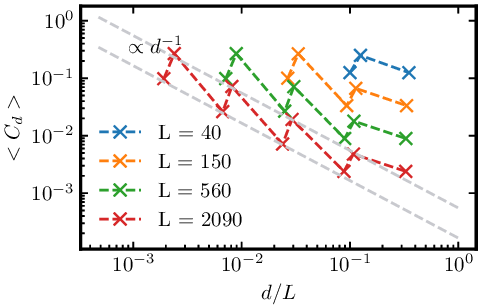}
    \caption{Correlation function $\left<C(d)\right>$ as a function of the distance $d/L$ around the system border for different system sizes $L$. For each size only the non-zero values of the correlation, $C(d) \neq 0$, are shown. Dashed lines are asymptotic limits for the largest system size, proportional to $d^{-1}$ [Eq.~(\ref{eq:correlationAsymptotic})]}.
    \label{fig:correlationSizes}
\end{figure}

To further confirm the CFT nature of the boundary, we compute its two-point correlation function.  Here, any spin correlation $\langle S_i S_j \rangle$ vanishes due to the subsystem symmetry. The symmetry-invariant  correlation is  $\langle E_{e_i} E_{e_j} \rangle = 1$ (resp. 0) if edges $e_i$ and $e_j$ are (resp. not) on the same geodesic. Hence, the correlation
\begin{equation}
\label{eq:Correlation}
\begin{split}
    C(d) &=\frac{1}{L}\sum_{i} \langle E_i E_{i+d} \rangle 
\end{split}
\end{equation}
between boundary points should be non-vanishing a priori. By computing the number of geodesics whose endpoints are separated by a distance $d$ along the boundary, we could numerically measure the correlation $C(d)$ as shown in Fig.~\ref{fig:correlationSizes}. Its zig-zag evolution is due to the presence of two different types of boundary polygons, which split the geodesics into two families, $X$ and $Y$. Expanding the polygon inflation rules of the hyperbolic lattice to a geodesic inflation, we derived an analytical asymptotic limit for the correlation of both types of geodesics,
\begin{equation}
C^{X}(d) = \frac{11}{2L} \frac{1}{d} \quad \&\quad
C^{Y}(d) = \frac{2}{\sqrt{3}L} \frac{1}{d},
\label{eq:correlationAsymptotic}
\end{equation}
decaying as a power law, as expected for a CFT \cite{francesco1997Bookconformal}, and agreeing quantitatively with numerics (see dashed lines in Fig.~\ref{fig:correlationSizes} for the largest system size).

Our numerical and analytical results thus establish that the boundary state of the HFM   exhibits features characteristic of a CFT at zero temperature in the continuum thermodynamic limit. With this in mind, we can now turn our attention to the main question of our paper, the consequence of  lattice defects in the HFM.

\begin{figure}[b]
    \centering
    \includegraphics[width = \linewidth]{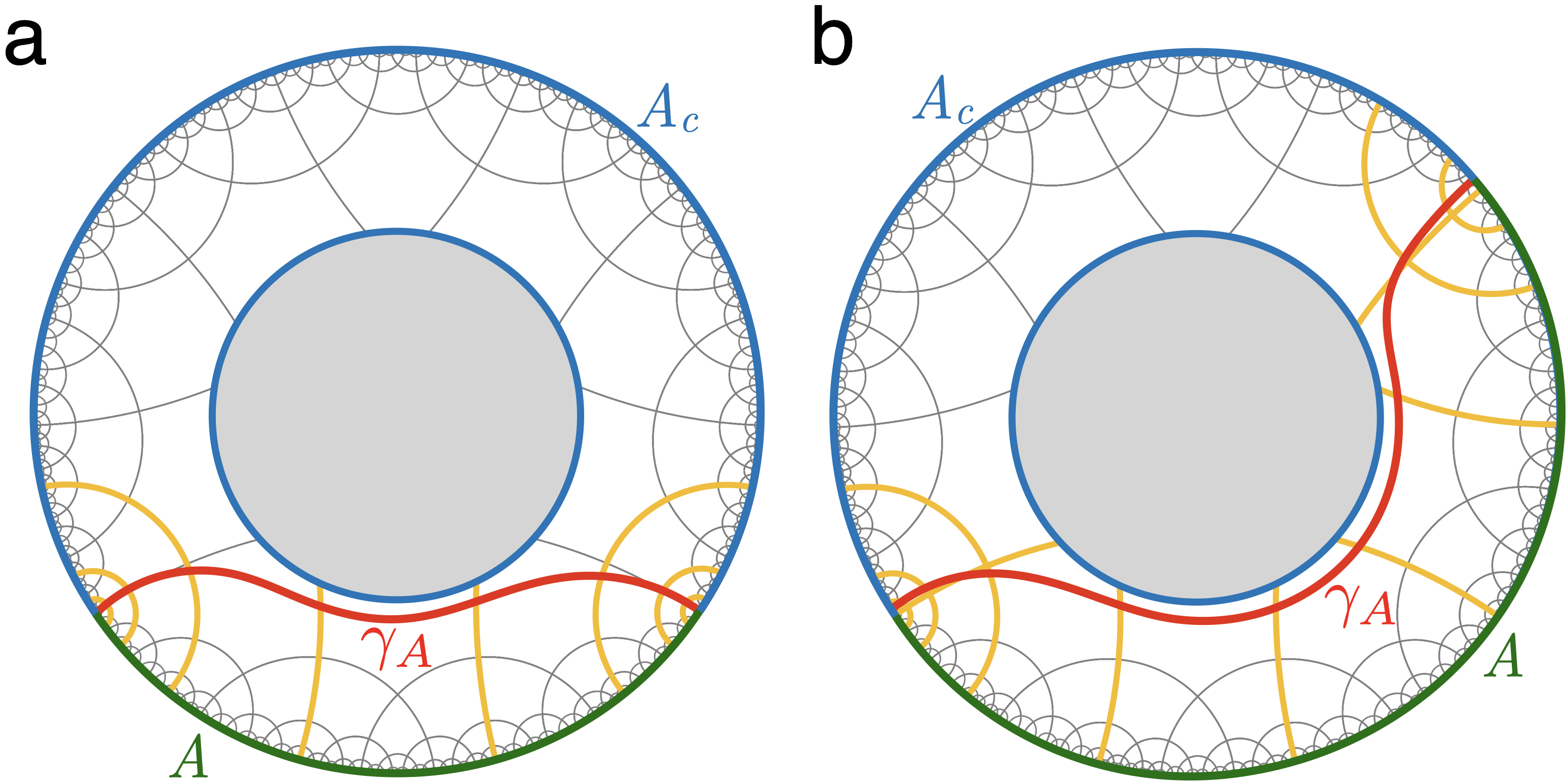}
    \caption{(a) Hyperbolic Fracton Model with a lattice defect, colored in gray. The geodesics connecting $A$ (colored in green) and $A_c$ (colored in blue,  includes the lattice defect boundary too)  are highlited in yellow and contribute to the entanglement entropy. The minimal covering curve $\gamma_A$ is higlighted in red.
    (b) The same figure for a larger region of $A$, showing how the minimal covering curve $\gamma_A$ changes.}
    \label{fig:FIG_3_EE_BH}
\end{figure}
 
\subsection{Lattice defects form a black hole}

The lattice defects are introduced by removing all spins and vertices within a circle of a given size at the center of the Poincar\'e disk [Fig.~\ref{fig:FIG_3_EE_BH}]. As a result, the ground state degeneracy increases due to the reduction in constraints \cite{Yan2019PhysRevBfracton2}. In the vertex model representation, the degeneracy increase is captured by the lattice geodesics separated into two segments by the defect. While arrows within each segment must remain aligned, the two segments themselves become uncorrelated. 
In this context, the complement region, $A_c$, is defined as the union of lattice boundary  excluding  region $A$,  and the boundary of the defect in the bulk. This useful picture enables to directly compute the entanglement entropy of the boundary state in presence of defects, by counting the geodesics  that connect  $A$ either to the remaining part of the lattice boundary or to the defect itself [Fig.~\ref{fig:FIG_3_EE_BH}]. From this perspective, it becomes clear that the minimal covering curve separating $A$ and $A_c$ qualitatively resembles the bulk minimal covering surface dual to the boundary entanglement entropy in 
AdS/CFT, in the presence of a black hole~\cite{Ryu2006PhysRevLett,Ryu2006JHEP}. See Fig.~\ref{fig:FIG_3_EE_BH} for the illustration. 

\begin{figure}
    \centering
    \includegraphics[width = \linewidth]{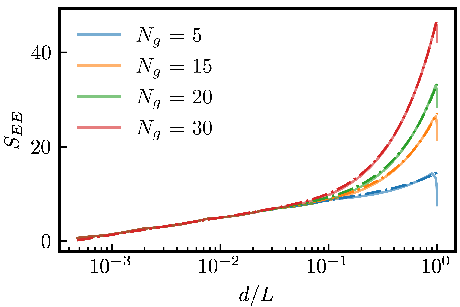}
    \caption{
    Entanglement entropy $S_{EE}$ as a a function of boundary segment $d$ for different defect sizes. The defect size is measured as a the number of geodesics $N_g$ that cross its boundary. Each curve follows closely the Calabrese-Cardy equation [Eq.~(\ref{eq:Calabrese-Cardy-Temp})] for the entanglement entropy of an infinite system at finite temperature for $d/L\lesssim 1$, shown in dashed line. The system size is $L=2800$.
    }
    \label{fig:entropyVsBlackHoleSize}
\end{figure}

\begin{figure}
    \centering
    \includegraphics[width = \linewidth]{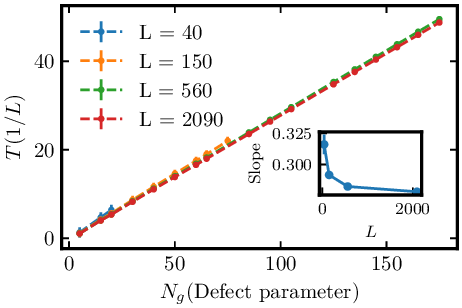}
    \caption{Temperature $T=1/\beta$ obtained by fitting the curves of Fig.~\ref{fig:entropyVsBlackHoleSize} to the Calabrese-Cardy equation as a function of defect perimeter length measured in the number of geodesics $N_g$ crossing it. We obtain a linear relation whose slope, depicted in the inset, approaches an asymptotic value with increasing system size $L$.
    }
    \label{fig:Temperature-linear}
\end{figure}

We find that the entanglement entropy computed for different number of defects, quantitatively fit  the Calabrese-Cardy equation of an infinite $(1+1)$-dimensional CFT at a finite inverse temperature $\beta$ given by \cite{calabrese2004entanglement}  [Fig.~\ref{fig:entropyVsBlackHoleSize}], 
\begin{equation}
S_{EE} = \frac{c}{3} \ln \left( \frac{\beta}{\pi a} \sinh \left( \frac{\pi d}{\beta}\right)  \right)\ .
\label{eq:Calabrese-Cardy-Temp}
\end{equation}
Note that here,  the only free parameter in this expression is the inverse temperature $\beta$, since the values of $c$ and $a$ were obtained from fitting the zero temperature behavior in Fig.~\ref{fig:zeroTempEntropies}. This is a crucial result which allows us to unambiguously identify an emergent boundary temperature arising from the presence of a circular lattice defect. On a hyperbolic lattice, the perimeter of this circular lattice defect can be measured by the number  of geodesics it intersects, denoted $N_g$. Plotting the evolution of this emergent temperature $T=1/\beta$ as a function of the perimeter $N_g$ in Fig.~\ref{fig:Temperature-linear}, we find an essentially perfect linear scaling relation, whose slope converges to an asymptotic value in the thermodynamic limit (see inset).

This behavior mirrors the physics of non-rotating  Ba\~nados--Teitelboim--Zanelli (BTZ) black holes~\cite{banados1992black}, whose temperature is proportional to the horizon radius.  
This is our main result: we were able to construct an emergent black hole by simply cutting out  bulk pentagons and hiding all the spins and interactions associated with them. 
The open boundary created by the lattice defects in the HFM emulates a black hole horizon within the AdS/CFT framework,
and dictates the boundary state effective temperature.

\begin{figure}[h]
    \centering
    \includegraphics[width = \linewidth]{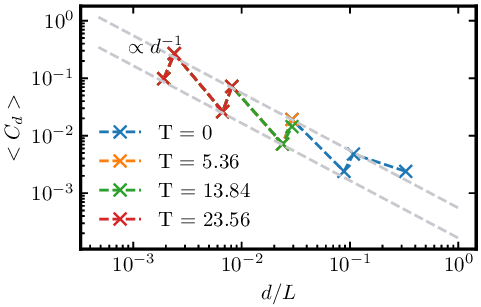}
    \caption{Correlation as a function of distance between boundary sites for different defects sizes labeled by the corresponding temperature. For each curve, only the non-zero values of the correlation are plotted. The system size is $L=2800$.}
    \label{fig:CorrelationsBH}
\end{figure}

Finally, we compute the correlation function \eqref{eq:Correlation} in presence of defects [Fig.~\ref{fig:CorrelationsBH}], whose behavior is consistent with the correlation function of a CFT at finite temperature. $C(d)$  decays as a power law at short distances $d \ll \beta$, and vanishes at long distances (comparing Figs.~\ref{fig:correlationSizes} and \ref{fig:CorrelationsBH}) because longer geodesics are cut by the black hole.

\section{Discussion}

In this work, we explored the boundary physics of the Hyperbolic Fracton Model (HFM) and demonstrated how it serves as a toy model for AdS/CFT duality. By introducing lattice defects into the bulk of the HFM, we show that these defects play an analogous role to black holes in the context of gravitational holography. More concretely, we reveal that the boundary entanglement entropy and correlation functions of the HFM exhibit the expected behavior of a conformal field theory (CFT) at finite temperature, with the temperature proportional to the black hole perimeter. Our work connects the well-known relationship between black hole in the bulk of AdS spacetime and finite-temperature CFTs on the boundary to fractonic matter with open boundaries and lattice defects, 
Our findings   align with recent developments in holographic tensor networks and the emergence of $p$-adic AdS/CFT in the HFM~\cite{Pastawski2015,Jahneaaw0092,yan2023arXivpadic}.
But equipped with a   bulk Hamiltonian, our model goes beyond holographic tensor-networks which only define   boundary states or  mappings between bulk and boundary Hilbert space.  This establishes the HFM as a fertile ground for exploring key aspects of gravitational holography and black hole information.

Looking forward, these results open up several directions for further exploration. The connection between lattice defects and emergent temperature could be further analyzed in more complex holographic setups, including multi-defect configurations and other tessellations and subsystem symmetries. Additionally, understanding the full range of defect-induced phenomena in HFM could shed light on broader aspects of AdS/CFT and its extensions, such as higher-dimensional generalizations or the incorporation of dynamical bulk geometry. Finally, the parallels with holographic tensor networks suggest potential experimental realizations of these concepts in quantum simulation platforms and electrical circuits \cite{chen23a,Dey24a}, providing an exciting opportunity to probe gravitational holography in controlled, lab-based systems.

\section*{Methods}

\textbf{Entanglement entropy.} To compute the entanglement entropy, we used the Python package HyperLattice ~\cite{schrauth2024hypertiling} to generate a $\left\{ 5,4 \right\}$ hyperbolic lattice with a specified number of layers, $ l $. This package provides a list of all vertices in the lattice along with their positions in the Poincar\'e disk, which allows us to compute a list, $ \Gamma $, of all pentagonal geodesics in the lattice. In particular, for each geodesic $\gamma$, we determined the positions of its endpoints and the number of border pentagons, $d_{\gamma}$, between its endpoints. As shown in Ref. ~\cite{Yan2019PhysRevBfracton1}, the entanglement entropy for a connected boundary segment $ A $ of size $ d $ is proportional to the number of lattice geodesics $N_{A,A_c}$ that connect $ A $ to its complement $ A_c $. We calculated $N_{A,A_c}$ by counting, for all possible boundary segments $ A $ of size $ d $, the number of pentagonal geodesics $ \gamma \in \Gamma $ that have one endpoint in $ A $ and the other in $ A_c $.

\textbf{Correlation.} To numerically compute the correlation $C(d)$ we sorted the geodesics $\gamma \in \Gamma$ by their length $d_{\gamma}$ and counted the number of geodesics for each possible length $d$. The correlation was also computed analytically expanding the polygon inflation procedure used in \cite{saraidaris2024critical,yan2023p} incorporating geodesic inflation rules. This approach relies on distinguishing two types of border geodesics, $X$ and $Y$, thus computing the border length $d$ between the endpoints of each type of geodesic and the amount of geodesics, which corresponds to the correlation, as a function of $d$ for each $L$. In the asymptotic limit for large $L$ the correlation follows the expected $1/d$ decay for a CFT
\begin{align}
        &C^{X}(d) = \frac{11}{2L} \frac{1}{d} \\ 
        &C^{Y}(d) = \frac{2}{\sqrt{3}L} \frac{1}{d}
\end{align}
with $d$ the distance between the endpoints of the geodesic in units of the border length $L$. The two different proportionality factors for each type of geodesic explains the zig-zag like pattern found in Fig. \ref{fig:correlationSizes}.

\textbf{Correlation and entanglement entropy with defects.} When a defect is introduced in the lattice, some of the geodesics in $\Gamma$ cross this defect and are split in two. This, effectively, increments the number of pentagon geodesics belonging to the list $\Gamma$, forming a new list $\tilde{\Gamma}$. The correlation and entanglement entropy were then computed as in the case with no defect using $\tilde{\Gamma}$.

The code implementation and data are available at \cite{FractonGitHub}.

\section*{Acknowledgements}
H.Y. thanks Alexander Jahn, Charles Cao, and Qiang Wen
 for helpful discussions.
H.Y. is supported by the 2024 Toyota Riken Scholar Program from the Toyota Physical 
and Chemical Research Institute, and the  Grant-in-Aid for Research Activity Start-up from Japan Society
for the Promotion of Science (Grant No. 24K22856). L.D.C.J.\ acknowledges financial support from ANR-23-CE30-0038-01 and from the CNRS International Research Project COQSYS, and thanks IFLYSIB (CONICET-UNLP) for their hospitality during this project. A. C. D. and M. S. were supported by Agencia I+D+i (Argentina) PICT2020/00520, CONICET (Argentina)
PIP2022/11220210100731CO and Universidad Nacional
de La Plata (Argentina) UNLP 11/X787.

\bibliography{cft_hfm_ref.bib}

\end{document}